# Enhanced Born Charge and Proximity to Ferroelectricity in Thallium halides

Mao-Hua Du and David J. Singh

Materials Science and Technology Division and Center for Radiation Detection Materials and Systems, Oak Ridge National Laboratory, Oak Ridge, TN 37831, USA

Electronic structure and lattice dynamics calculations on thallium halides show that the Born effective charges in these compounds are more than twice larger than the nominal ionic charges. This is a result of cross-band-gap hybridization between Tl-$p$ and halogen-$p$ states. The large Born charges cause giant splitting between longitudinal and transverse optic phonon modes, bringing the lattice close to ferroelectric instability. Our calculations indeed show spontaneous lattice polarization upon lattice expansion starting at 2%. It is remarkable that the apparently ionic thallium halides with a simple cubic CsCl structure and large differences in electronegativity between cations and anions can be very close to ferroelectricity. This can lead to effective screening of defects and impurities that would otherwise be strong carrier traps and may therefore contribute to the relatively good carrier transport properties in TlBr radiation detectors.

# I. Introduction

Room temperature radiation detection is an important technology that finds applications in many areas such as medical imaging, nuclear safeguard and national security. Many semiconductor materials have been investigated for their radiation detection capabilities.[1] The performance of the semiconductor radiation detectors relies on the collection efficiency of free carriers generated by radiation energy deposition. Several important criteria have been established for material selection, i.e., (1) large atomic number for high gamma-ray stopping efficiency, (2) high resistivity (that reduces dark current and device noise), (3) large enough band gap of >1.4 eV (suitable for room temperature applications), and (4) large μτ product (μ and τ are carrier mobility and lifetime, respectively).[2]

TlBr has been investigated for radiation detection for several decades.[3,4,5] Some very encouraging results have been obtained,[6,7,8,9,10,11,12] showing that TlBr is a promising material for room temperature spectroscopic gamma-ray detection. TlBr has high atomic numbers (Tl: 81 and Br: 35) and high density (7.56 g/cm$^3$), which lead to efficient radiation absorption. The high resistivity of TlBr (~$10^{10}$ ohm-cm) reduces the dark current. The band gap of 2.68 eV enables room-temperature applications. The advancement of the material growth and purification techniques in recent years has significantly increased the μτ product to ~$10^{-3}$ cm$^2$/V for electrons and ~$10^{-4}$ cm$^2$/V for holes.[6,7,8,9,10] These μτ results are comparable to those for CdZnTe (CZT), which is the current state-of-the-art room-temperature semiconductor radiation detector material.[1]

It is worth noting that when one moves from III-V to II-VI and I-VII compound semiconductors, the carrier mobility tends to decrease significantly. High electron mobility is usually found in strongly covalently bonded semiconductors, such as GaAs. TlBr has simple cubic CsCl structure [Space group 221, Tl (0.0, 0.0, 0.0) and Br (0.5, 0.5, 0.5)], a wide band gap of 2.68 eV, and a large difference in electronegativity between Tl and Br (Tl: 1.62; Br: 2.96), all of which suggest relatively high ionicity. The more ionic I-VII materials also tend to have softer lattices, which are prone to structural defects. Therefore, the high $\mu\tau$ achieved in TlBr is rather remarkable. This is particular so considering the lower carrier mobility of TlBr than that of CZT,[1] which means that the large $\mu\tau$ product corresponds to a very long carrier lifetime in TlBr. The carrier lifetime is closely related to defects in the material, which are in general complicated, varying from one material to another and depending on growth conditions and sample purity. Here, we investigate some fundamental properties of TlBr in order to identify features that allow TlBr to have a large $\mu\tau$.

By scanning basic physical properties of many semiconductor radiation detector materials, we find that one feature that stands out in TlBr is its high static dielectric constant ($\varepsilon_{st}$) of ~30.60 (290 K). Similar to TlBr, TlCl also has the CsCl structure and a high $\varepsilon_{st}$ of 32.70 (293 K).[13] In comparison, other semiconductor compounds that have been investigated for radiation detection have much lower static dielectric constants, e.g., 10.9 (CdTe), 10 ($Cd_{0.9}Zn_{0.1}Te$), 10.2 (CdSe), 12.8 (GaAs), 12.4 (InP), 11.7 (Si), 16 (Ge), 9.7 (4H-SiC), 8.8 ($HgI_2$).[1, 13] The dielectric constant usually decreases with increasing band gap and ionicity. The exceptional high dielectric constants for TlCl and TlBr are mainly due to the large ionic (lattice) polarization under the macroscopic electric field. The electronic contribution to the static dielectric constant, i.e., the optical dielectric constant ($\varepsilon_{opt}$), is only 5.34 in TlBr consistent with the expected trend with band gap, while its $\varepsilon_{st}$ is 30.6 (both measured at 290 K).[13][14] Similarly, $\varepsilon_{opt}$ (4.76) in TlCl is much smaller than $\varepsilon_{st}$ (32.7) (both measured at 293 K).[13] The large difference between $\varepsilon_{st}$ and $\varepsilon_{opt}$ in TlCl and TlBr are consistent with the observed giant splitting of the longitudinal and transverse optical phonons (LO and TO),[15][16] satisfying Lyddane-Sachs-Teller relationship ($\omega_{LO}/\omega_{TO} = \sqrt{\varepsilon_{st}/\varepsilon_{opt}}$). A large dielectric constant should lead to more effective screening of charged defects and impurities that can scatter or trap free carriers, leading to improved carrier mobility and carrier lifetime.

Here we show enhanced Born effective charges for thallium halides (which are more than twice larger than their nominal ionic charges) based on density functional perturbation theory. We will show the electronic structure of TlBr exhibits cross-bandgap hybridization between the Br-$p$ and Tl-$p$ states, which leads to the large Born charges. The resulting large LO – TO splitting places these materials near ferroelectric instability. According to our calculations, expanding the lattice constants of TlCl, TlBr, and TlI by 2% will lead to spontaneous lattice polarizations. The large enhancement of Born charges and the proximity to spontaneous lattice polarization, which give rise to large dielectric constant, may be an useful search criterion for the semiconductor radiation detector materials among the relatively soft ionic compounds.

## II. Methods

We performed density functional calculations to study the electronic structure and lattice dynamics of thallium halides. Two generalized gradient approximations (GGA) were used. Standard Perdew-Burke-Ernzerhof (PBE)[17] exchange-correlation functionals were used for total energy and lattice dynamics calculations, while the Engel-Vosko GGA was used for electronic structure calculations since this functional was designed to reproduce as well as possible the exact exchange correlation potential rather than the total energy, and as a result gives significantly improved results such as band gap and electronic dielectric functions may be expected.[18][19][20]

The electronic structures were calculated using general potential linearized augmented planewave (LAPW) method, as implemented in the wien2k package.[21] We used LAPW sphere radii for 2.9 Bohr for both Tl and Br with converged Brillouin zone samplings and basis sets, including local orbitals. Spin-orbit coupling was included using a second variational approach. The lattice dynamics calculations were performed using linear response in the quantum espresso code[22] with norm-conserving pseudopotentials and the Perdew-Burke-Ernzerhof (PBE)[23] exchange-correlation functionals. A cutoff energy of 150 Ry was used.

## III. Results and Discussion

The band structure is shown in Fig. 1 and the corresponding electronic density of states (DOS) and projections onto the LAPW spheres is showm in Fig. 2. These calculations were done using the experimental lattice constant of 3.9842 Å.[24] As may be seen, an indirect band gap of 2.0 eV is obtained. Although the Engel-Vosko GGA generally gives larger band gaps in better agreement with experiment than other standard functionals, such as the local density approximation, this is still underestimated compared to the experimental band gap of 2.68 eV. The general features of the band dispersion are similar to previous calculations.[25] The four valence bands shown come from the Tl 6$s$ and Br 4$p$ orbitals while the conduction bands are mainly Tl-$p$ states. Although the Tl-$s$ and Br-$p$ bands are hybridized with each other, the lowest band in Fig. 1 has predominant Tl $s$ character, while the Br $p$ character is concentrated between -3 eV and 0 eV with respect to the valence band maximum (VBM). The strong hybridization between the Tl-$s$ and Br-$p$ bands results in relatively large band dispersion and large Tl-$s$ component near the VBM. There is also considerable cross-gap hybridization between the Br $p$ and nominally unoccuppied Tl 6$p$ states, reminiscent of the cross-gap hybridization that is important in ferroelectric oxides, such as $PbTiO_3$.[26,27,28,29] In those materials, the hybridization leads to enhanced Born effective charges, very large LO-TO splitting, and ferroelectricity when the spitting is large enough to drive infrared active transverse optic modes unstable.

Before turning to the lattice properties, we briefly discuss the optical properties from the electronic structure calculation. The calculated optical dielectric function is shown Fig. 3. This was calculated for vertical electric dipole transitions, using the optical package of the wien2k code. As may be seen, the real part of the dielectric function at low frequency is $\varepsilon_{opt}$ = 5.2, in good agreement with the experimental value, which varies from 5.64 to 5.45 in the temperature range of 2-290 K.[13] This is much lower than the measured static dielectric constant, which means that the lattice part is dominant in the static dielectric constant. This is consistent with experimental measurements, as discussed above.

Phonon dispersions for TlCl, TlBr, and TlI are shown in Fig. 4. These calculations were done by using experimental lattice constants (TlCl: 3.83 Å[16]; TlBr: 3.9842 Å[24]; TlI: 4.205 Å[30]) for simple cubic CsCl structures. Among these compounds, TlCl and TlBr crystallize in the CsCl structure while TlI crystallizes in orthorhombic structure but transforms to the high temperature CsCl structure with both increasing temperature and pressure.[31] For all three compounds, we observe soft phonons and large LO-TO splitting (see Table I). Our calculated phonon dispersions are in good agreement with available neutron scattering data.[15][16] The softening near the M point in TlCl [see Fig. 4(a)], which is close to lattice instability, also agrees well with the experimental results. The large LO-TO splittings result from large Born effective charges, which measure how lattice polarization develops with atomic displacement. The calculated Born effective charges for ions in all three compounds (see Table I) are more than twice larger than their nominal ionic charges (Tl: +1 and halogen: -1). This is due to the cross-gap hybridization between the halogen-$p$ and the spatially extended Tl-$p$ states (as can be seen in Fig. 2 for TlBr), which results in charge transfer between the nearest-neighbor Tl and halogen ions upon atomic displacement. The anomalously large Born effective charges reflect significantly enhanced dynamic coupling between atomic displacement and polarization.

The cross-gap hybridization and the resulting enhancement of Born effective charges are seen in ferroelectric oxides, such as $PbTiO_3$.[26][27][28][29] This suggests that thallium halides may be close to ferroelectric instability. Indeed, we find spontaneous lattice polarization for all three thallium halides upon expansion of lattice constant by 2%. We did calculations for displacements along [001], [011] and [111] directions. The ferroelectric instability is strongest for the displacement along the [001] direction. This is the direction of the square face of the cubic cages around the eight-fold coordinated ions in the CsCl structure. This corresponds to a tetragonal ferroelectric state. Fig. 5 shows that relative displacement of the Tl and Br sublattices along the [001] lattice direction becomes energetically favorable when the lattice constant is expanded by more than 2% relative to the experimental value. We also did some calculations within the local density approximation. We found similar results, except that the onset of ferroelectricity was at ~1% lattice expansion relative to experiment. Such spontaneous polarization is generic to the thallium halides in CsCl structure as shown in Fig. 5. In these cases, the long range Coulomb interaction overcomes the local lattice stabilization forces, such as the repulsion between the ionic shells of cations and anions, causing the lattice polarization. Cross-gap hybridization between halogen $p$ states and the spatially extended Tl $6p$ states plays a critical role in reducing the shell repulsion between cations and anions by enhancing bond covalency. It should be noted that spontaneous polarization is generally expected when sufficient lattice expansion occurs in an ionic crystal thereby reducing the ionic shell repulsion. What is significant is that the enhanced Born effective charges that we find bring thallium halides very close to such lattice instability.

Nearness to ferroelectricity due to large Born effective charges is well known in complex oxides, e.g., perovskites. However, it is remarkable that the apparently ionic thallium halides with a simple cubic CsCl structure and large differences in electronegativity between cations and anions (Tl: 1.62; Cl: 3.16; Br: 2.96; I: 2.66) can be very close to ferroelectricity and become ferroelectric with merely 2% expansion of the lattice constant. This adds these halides to the list of simple materials that are ferroelectric or near ferroelectricity, including group IV tellurides with NaCl structure (e.g. GeTe, SnTe), and according to recent calculations some NaCl structure binary oxides.[32] According to our results, actual ferroelectricity may be realized in strained hetero-epitaxial growth of thallium halide thin films.

Ionic and soft-lattice compounds often have inferior transport properties compared to covalent semiconducting compounds. For example, NaCl is far inferior to the same row group IV semiconductor, Si. Lattice defects in NaCl serve as strong and effective carrier traps. The high $\mu\tau$ product observed in TlBr (comparable to that of CZT) may therefore appear surprising. We suggest that the high $\mu\tau$ product may in part be due to more effective screening of defects and impurities due to the fact that TlBr has a large static dielectric constant. Obviously, the carrier mobility is improved by better screening of the charged defects and impurities that cause scattering. Also, better screening of the charged carrier traps leads to the reduction of carrier trapping cross section and consequently extended carrier lifetime. Thus, the soft-lattice semiconductor compounds close to ferroelectric instability may be surprising but valid candidates for the development of new semiconductor radiation detector materials.

## IV. Summary

We have calculated the electronic structure and lattice dynamics of thallium halides [Tl$X$ ($X$ = Cl, Br, I)]. The results show that the Born effective charges are more than twice larger than the nominal ionic charges, which are consistent with the giant LO-TO splitting found in these compounds. The large Born effective charges are explained by the cross-gap hybridization between Tl-$p$ and Br-$p$ states. This is the same mechanism as in many ferroelectric oxides and in the present case arises as a result of the low lying spatially extended 6p orbital on the Tl$^+$ ions. Indeed, our calculations show spontaneous lattice polarization in all three compounds when the lattice constant is expanded by more than 2%. The large Born effective charges in thallium halides are responsible for the dominant lattice contribution in the static dielectric constant. The large dielectric constant and the resulted effective screening of defects and impurities may contribute to the relatively good carrier transport properties in TlBr, whose μτ product is among the best in room-temperature semiconductor radiation detector materials. The insights gained in this study may be useful for understanding and improving the performance of TlBr radiation detectors and may open new opportunities for soft-lattice semiconductor compounds, which exhibit enhanced Born effective charges and proximity to ferroelectric instability, in the radiation detection applications.

## Acknowledgments

We are grateful for the helpful discussions with Zane W. Bell and Lynn A. Boatner. This work was supported by the U.S. DOE Office of Nonproliferation Research and Development NA22.

**TABLE I. Longitudinal and transverse optical phonon modes at Γ point ($\omega_{LO}$ and $\omega_{TO}$) and Born effective charges ($Z^*$) for TlCl, TlBr, and TlI.**

|  | TlCl | TlBr | TlI |
|---|---|---|---|
| $\omega_{LO}$ (cm$^{-1}$) | 51.8 | 38.5 | 31.3 |
| $\omega_{TO}$ (cm$^{-1}$) | 158.1 | 106.8 | 82.8 |
| $Z^*$ | 2.02 (Tl) | 2.10 (Tl) | 2.21 (Tl) |
|  | -2.02 (Cl) | -2.10 (Br) | -2.21 (I) |

FIGURE 1. Calculated band structure of TlBr, including spin orbit. Note the indirect band gap.

FIGURE 2. (Color online) Electronic DOS and projections onto the LAPW spheres. Note the cross-gap hybrization between the Br *p* bands and the nominally unnoccupied Tl 6*p* bands.

FIGURE 3. (Color online) Calculated optical dielectric function of TlBr.

FIGURE 4. (Color online) Phonon dispersions for (a) TlCl (b) TlBr and (c) TlI in simple cubic CsCl structure. Note the large LO-TO splitting in all three compounds.

FIGURE 5. (Color online) Energy changes as functions of displacements of the Tl ion in the [100] lattice direction of (a) TlCl, (b) TlBr, and (c) TlI for selected lattice constants from shrinking -2% to expanding 3% of the experimental lattice constant.

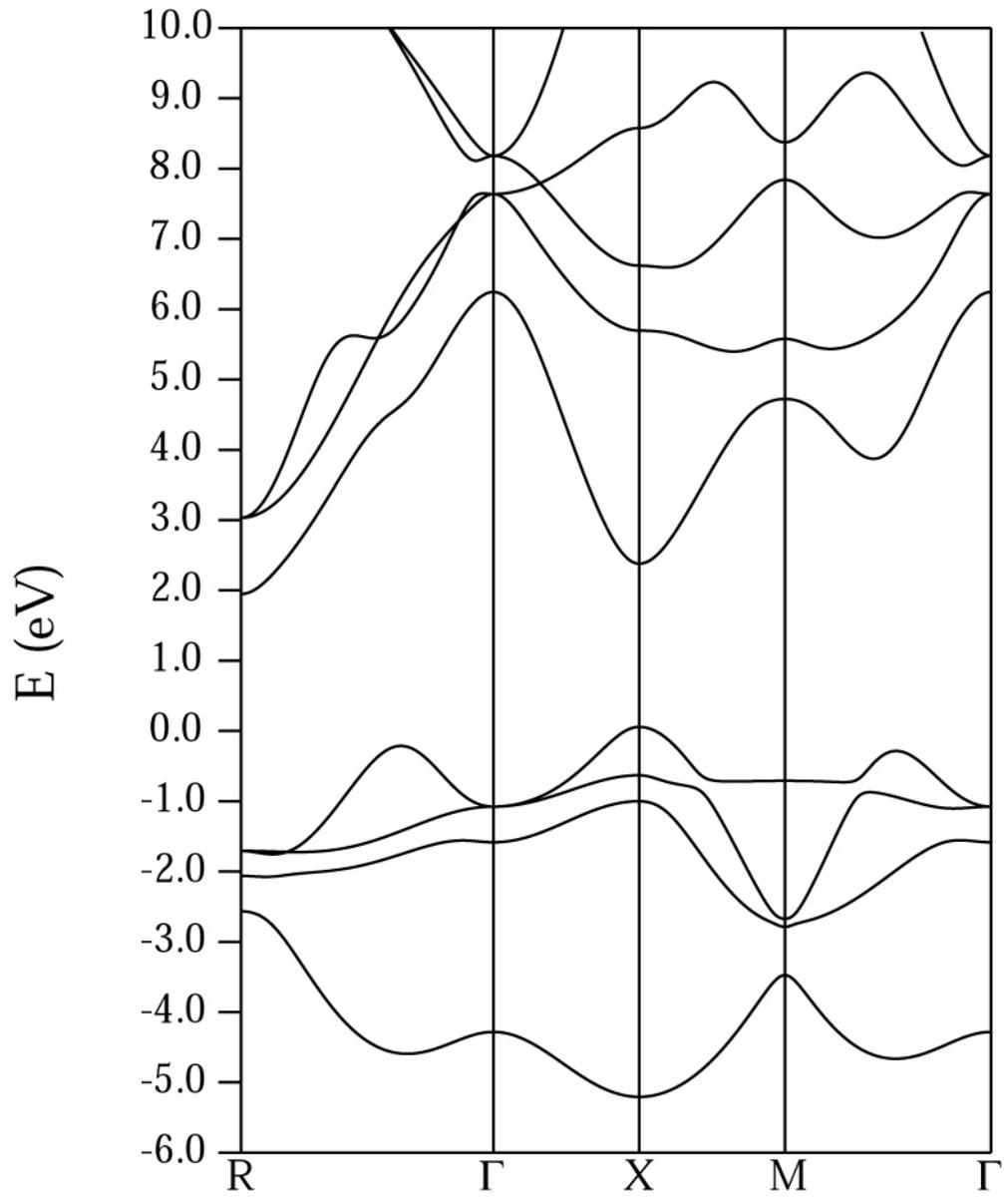

**FIG. 1**

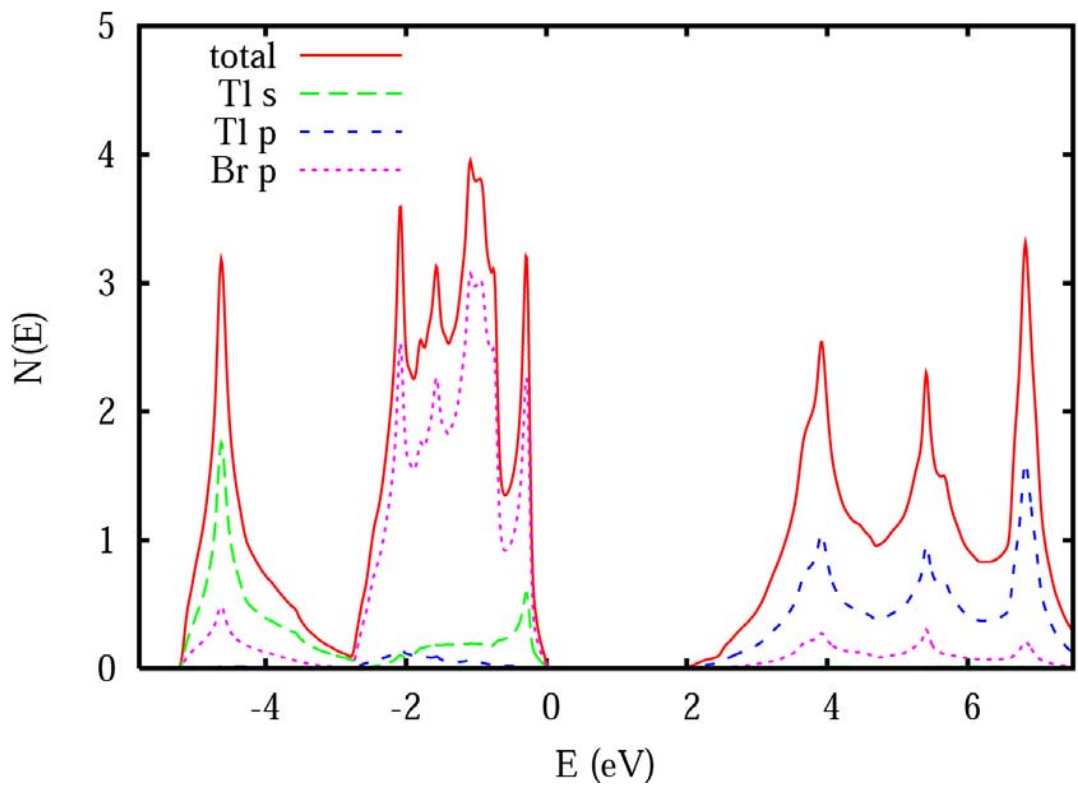

**FIG. 2**

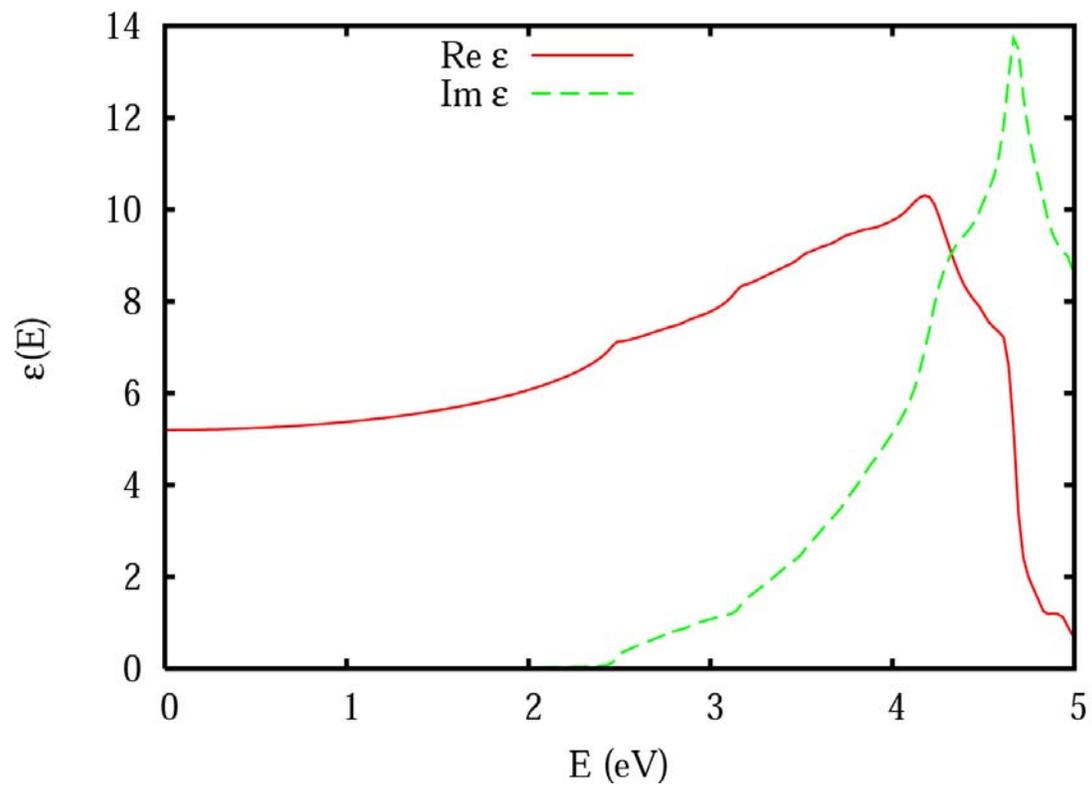

**FIG. 3**

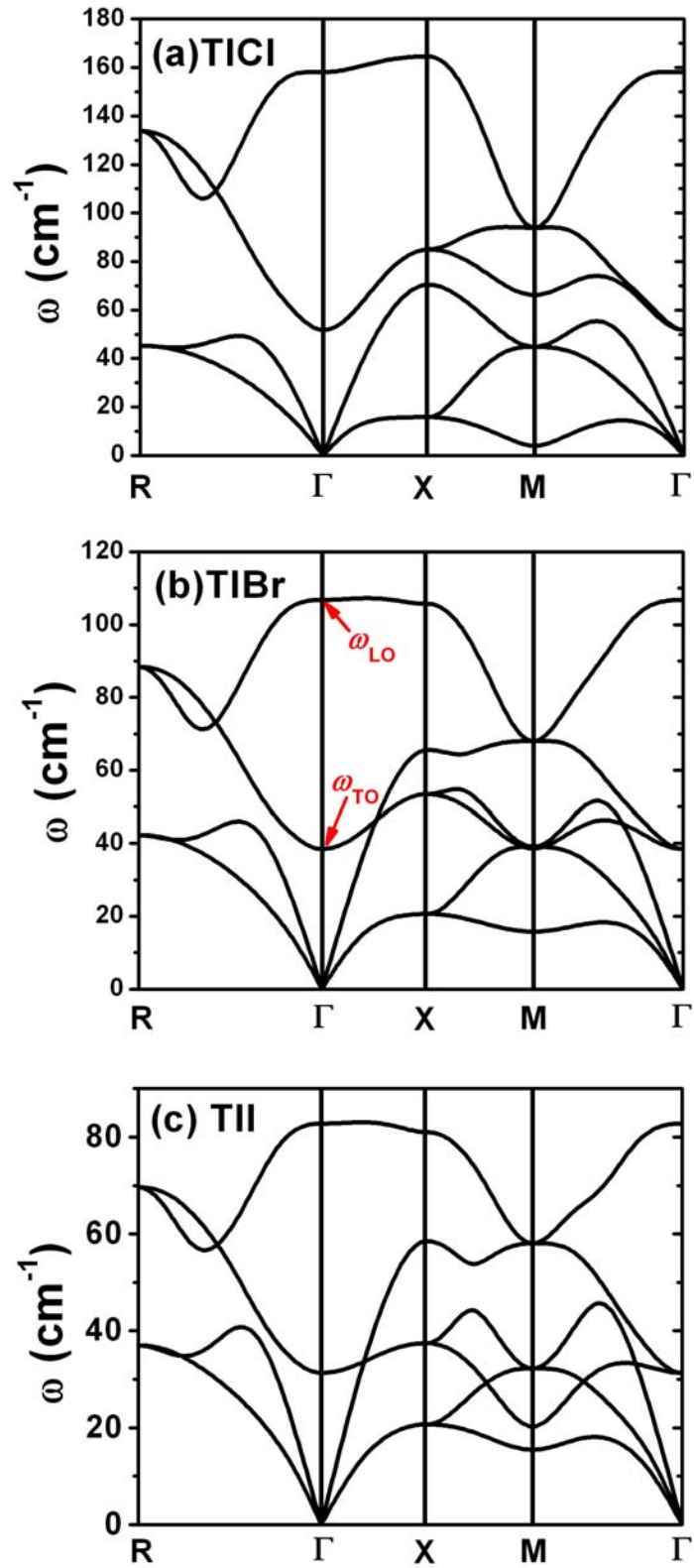

**FIG. 4**

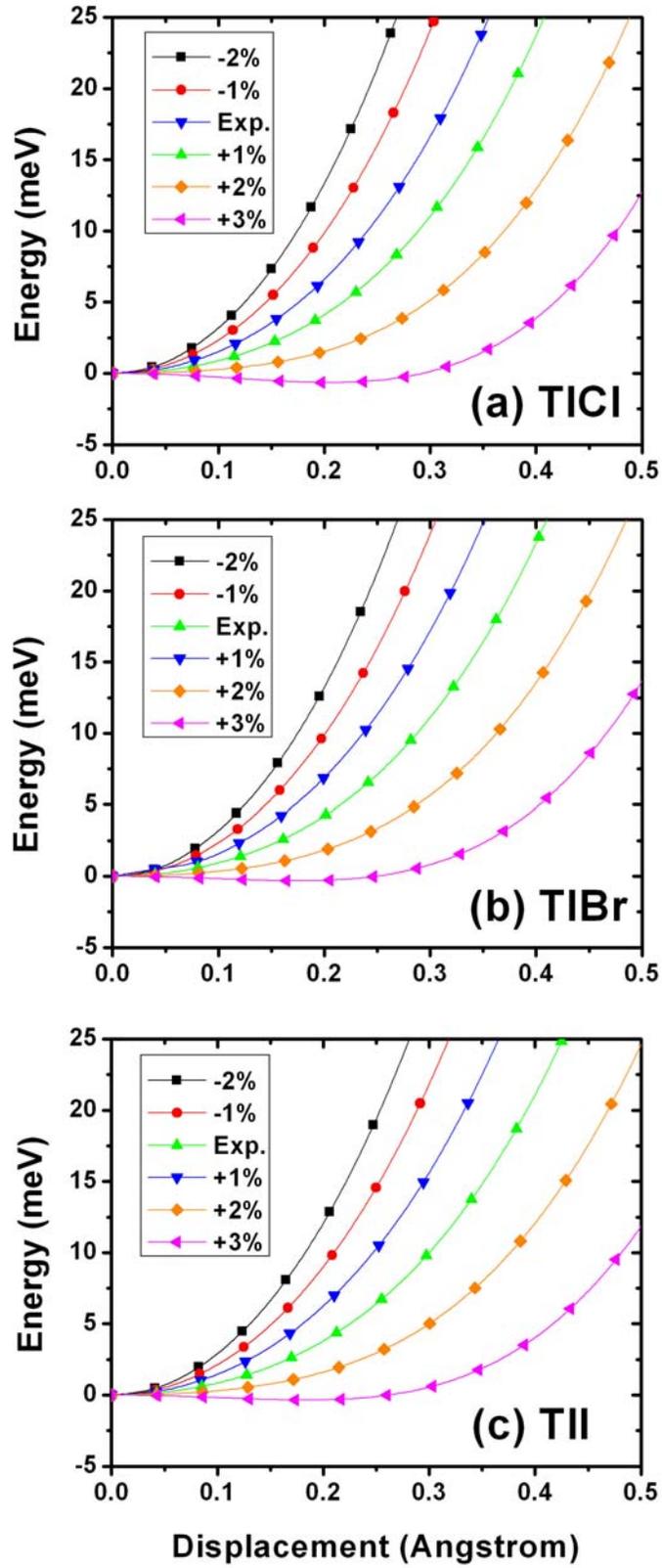

**FIG. 5**